\documentclass[a4paper]{jpconf}
\usepackage{iopams}
\usepackage{graphicx}
\usepackage{epstopdf}
\usepackage{amsmath,amssymb}

\newcommand{\pT}           {\ensuremath{p_{\rm T}}}
\newcommand{\pp}           {pp}
\newcommand{\PbPb}         {\mbox{Pb--Pb}}
\newcommand{\pPb}          {\mbox{p--Pb}}

\newcommand{\Npart}        {\ensuremath{N_\mathrm{part}}}
\newcommand{\Nparttar}     {\ensuremath{N_\mathrm{part}^\mathrm{target}}}

\newcommand{\Ncoll}        {\ensuremath{N_\mathrm{coll}}}

\newcommand{\qpa}          {\ensuremath{Q_{\rm pPb}}}

\newcommand{\Nch}          {\ensuremath{N_\mathrm{ch}}}

\begin{document}
\title{Centrality Dependence of Particle Production in p--A collisions measured by ALICE}

\author{Alberica Toia (for the ALICE Collaboration)}
\address{Goethe University Frankfurt, GSI Darmstadt}
\ead{alberica.toia@cern.ch}

\begin{abstract}
We present the centrality dependence of particle production in \pPb\
collisions at $\sqrt{s_{NN}}$ = 5.02 TeV measured by the ALICE experiment,
including the pseudo-rapidity and transverse momentum spectra, with a
special emphasis on the event classification in centrality classes and
its implications for the interpretation of the nuclear effects.
\end{abstract}

\section{Introduction}
Nuclear modification factors measured by ALICE in minimum bias (MB)
\pPb\ collisions for charged particles~\cite{alice_RpA_new}, heavy
flavor and jets show no deviations from unity at high-\pT, demonstrating that the observed strong suppression in
\PbPb\ collisions is due to final state effects. 
However several
measurements~\cite{Abelev:2012ola,Abelev:2013bla,Abelev:2013haa,ABELEV:2013wsa}
of particle production in the low and intermediate \pT\ region 
can not be explained by an incoherent superposition of
\pp\ collisions, but rather call for coherent and collective effects.
As their strength increases with multiplicity, a more detailed
characterisation of the collision geometry is needed. Moreover, a knowledge of the geometry dependence is necessary to interpret the suppression pattern of J/$\psi$ in \PbPb\ collisions relative to the effects observed in the nuclear medium produced in \pPb\ collisions.

ALICE has carried out detailed studies of the centrality determination
in \pPb\ collisions, and the possible biases induced by the event
selection on the scaling of hard processes in a selected event sample.
The centrality determination consists in relating a Glauber model,
which calculates the geometric properties of the event (\Ncoll), to a
measured observable related to the event activity in a specific
rapidity region~\cite{Alice:Centrality}, via the conditional
probability to observe a certain activity for a given \Ncoll.
Specifically, particle production measured by detectors at
mid-rapidity can be modeled with a negative binomial distribution
(NBD). The zero-degree energy is related to the number of the slow nucleons emitted
in the nucleus fragmentation process, which we model with a Slow
Nucleon Model (SNM)~\cite{Alice:CentralityPA}.  However, the connection of the
measurement to the collision geometry has to be validated, eg by
correlating observables from kinematic regions causally disconnected
after the collision, or by comparing a Glauber MC with data for a known
process, as eg the deuteron dissociation probability at RHIC~\cite{Adare:2013nff}.

In addition the consistency of the approach must be demonstrated.
As in general the selection in a system with large relative 
fluctuations can induce a bias, one needs to identify the physics origin of 
the bias in order to correct centrality dependent 
measurements.
In \pPb\ collisions, the relative large size of the multiplicity
fluctuations has the consequence that a centrality selection based on
multiplicity may select a biased sample of nucleon-nucleon
collisions. In essence, by selecting high (low) multiplicity one
chooses not only large (small) average \Npart\, but also positive
(negative) multiplicity fluctuations per nucleon-nucleon collision. These fluctuations are partly
related to qualitatively different types of collisions, described in
all recent Monte Carlo generators by impact parameter dependence of the number of
particle sources via multi-parton interaction.  However also other types of
biases have an influence on the nuclear modification factor: the
jet-veto effect, due to the trivial correlation between the centrality
estimator and the presence of a high-\pT\ particles in the event; the
geometric bias, resulting from the mean impact parameter between
nucleons rising for the most peripheral events.

\section{The ALICE approach}\label{sec:hybrid}  
The ALICE approach aims to a centrality
selection with minimal bias and, therefore, uses the signal in the Zero
Degree Calorimeter (ZNA). In this case we cannot establish a direct
connection to the collision geometry but we can study the correlation
of two or more observables that are causally disconnected after the
collision, e.g. because they are separated in rapidity.
\begin{figure}
\begin{center}
\includegraphics*[width=0.495\textwidth]{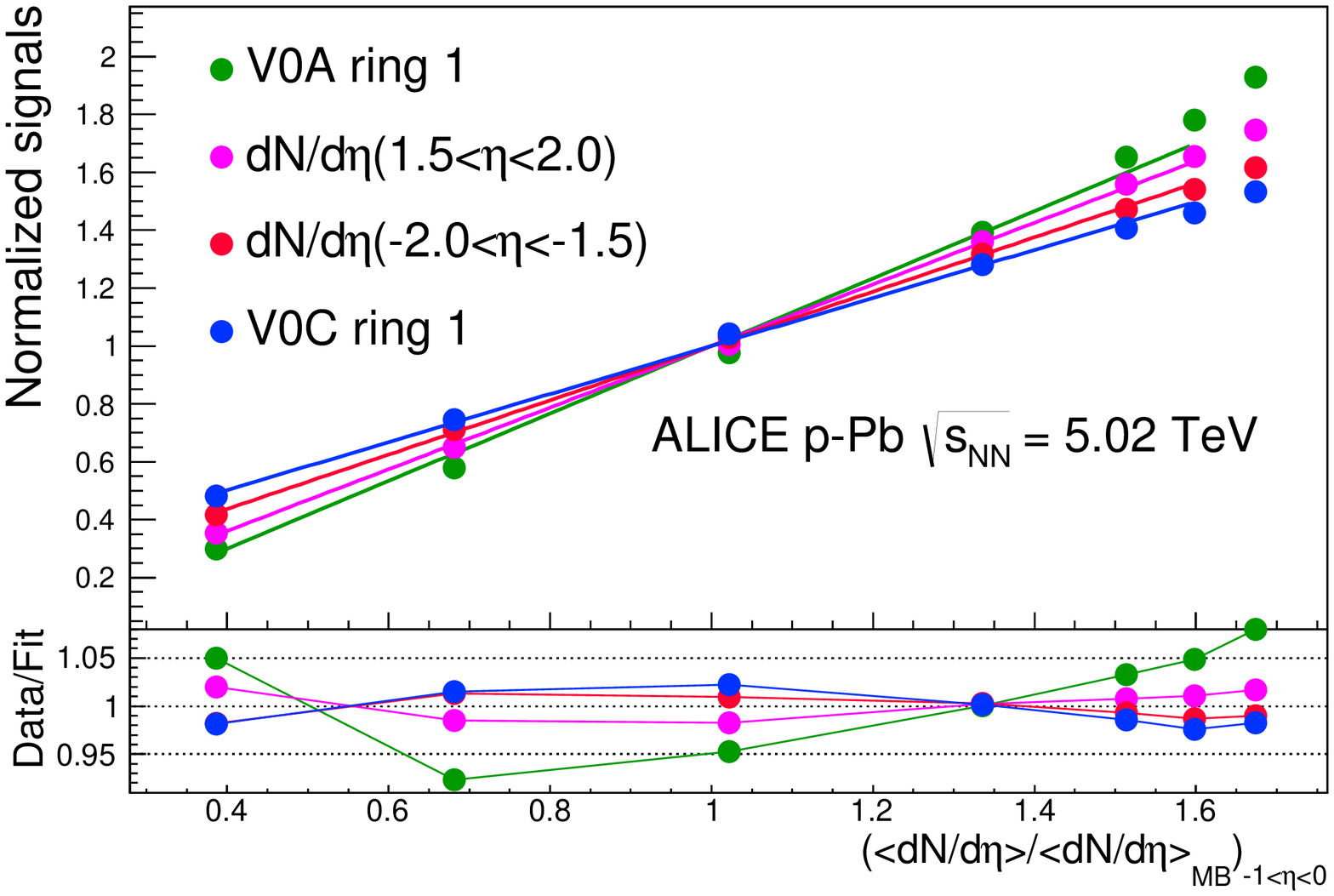}
\includegraphics*[width=0.35\textwidth]{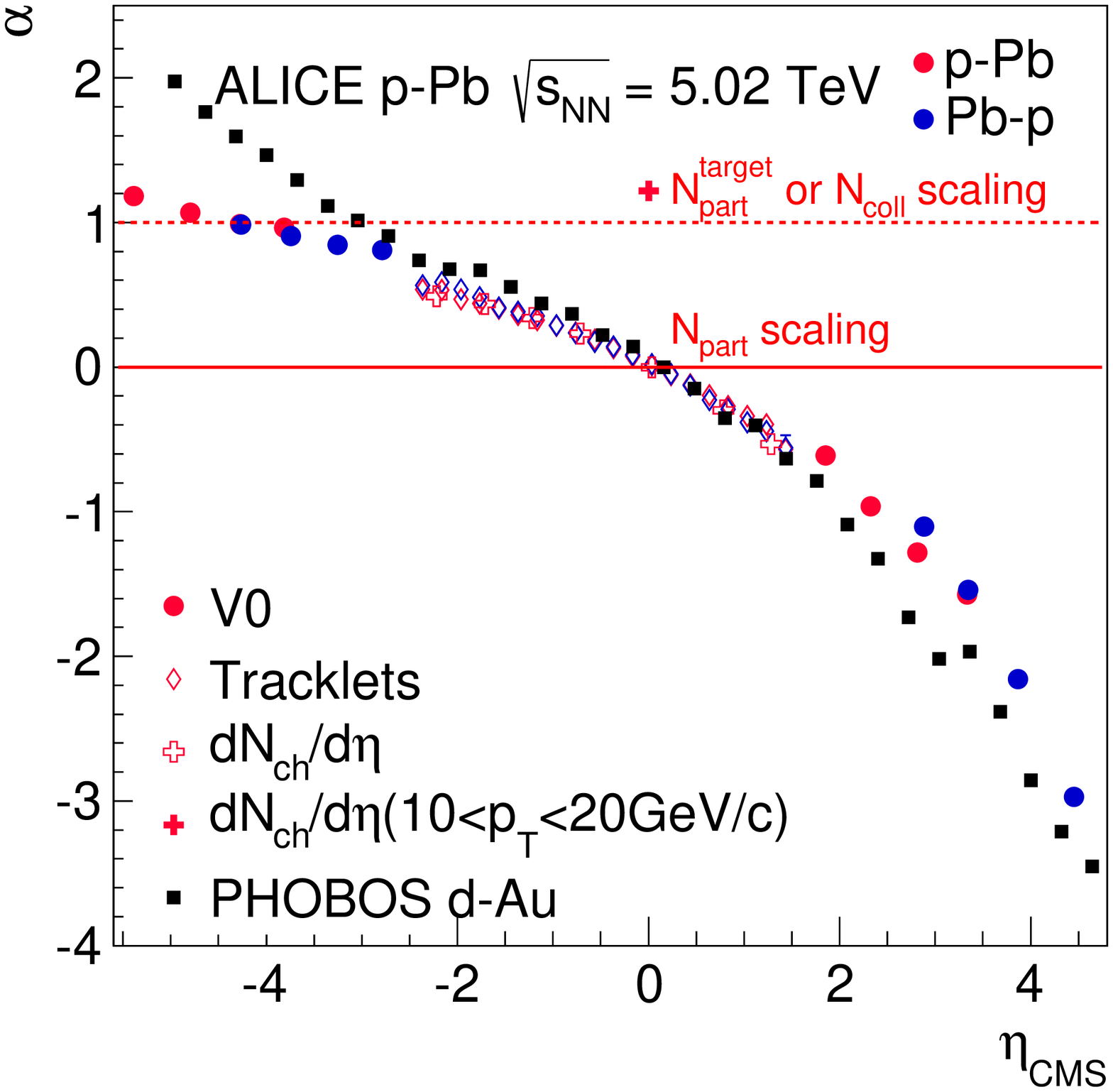}
\caption{Left: Normalized signal from various observables versus the
  normalized charged-particle density, fit with a linear function of
  \Npart. Right: Results from the fits as a function of the
  pseudorapidity covered by the observables. The red
  horizontal lines indicate the ideal geometrical
  scalings. The most central point is excluded from the fit, to avoid
  pile-up effects.}
\label{fig:hybrid}
\end{center}
\end{figure}
In centrality classes selected by ZNA, we study the dependence of
various observables in different $\eta$ and \pT\ regions on the
charged particle density at mid-rapidity.  Fig.\ref{fig:hybrid}
indicate a monotonic change of the scaling with rapidity.  The
correlation between the ZDC energy and any variable in the central
part shows unambiguously the connection of these observables to
geometry.  Exploiting the findings from the correlation analysis
described, we make use of observables that are expected to scale as a
linear function of \Ncoll\ or \Npart, to calculate \Ncoll:
\begin{itemize}
\item $\Ncoll^{\rm mult}$: the multiplicity at mid-rapidity
proportional to the \Npart;
\item $\Ncoll^{\rm Pb-side}$: the target-going multiplicity
proportional to \Nparttar;
\item $\Ncoll^{\rm high-\pT}$: the yield of high-\pT\ particles at
mid-rapidity is proportional to \Ncoll.  
\end{itemize}
These scalings are used as an ansatz, in the ALICE socalled
\emph{hybrid method} to calculate \Ncoll, rescaling the MB value
$\Ncoll^{\rm MB}$ by the ratio of the normalized signals to the MB
one.  We therefore obtain 3 sets of values of \Ncoll. The relative
difference does not exceed 10\%.  We have performed a consistency
check, correlating ZNA and V0A centrality measurements, to establish
their relation to the centrality.  The \Ncoll\ distributions for
centrality classes selected with ZNA, obtained from the SNM-Glauber
fit, are convolved with the NBD from the fit to the V0A
distribution. These are compared to the data and a good agreement is
found.  Moreover, a good agreement is also achieved with an unfolding
procedure, where the \Ncoll\ distributions have been fitted to the
data using the parameters from the NBD-Glauber fit.  For each V0A
distribution selected by ZNA, we find the \Ncoll\ distribution that,
convolved with the $\rm NBD_{\rm MB}$, fits the data, i.e. the
parameters of the fit are the relative contributions of each
\Ncoll\ bin.  The existence of \Ncoll\ distributions that folded with
NBD agree with measured signal distributions is a necessary condition
for ZNA to behave as an unbiased centrality selection This procedure,
which actually does not work for a biased centrality selection (eg
CL1) shows that the energy measured by ZN is connected to the
collision geometry.

\section{Physics Results}
\subsection{Nuclear Modification Factors}
\begin{figure}[t!f]
 \centering
\includegraphics[width=0.4\textwidth]{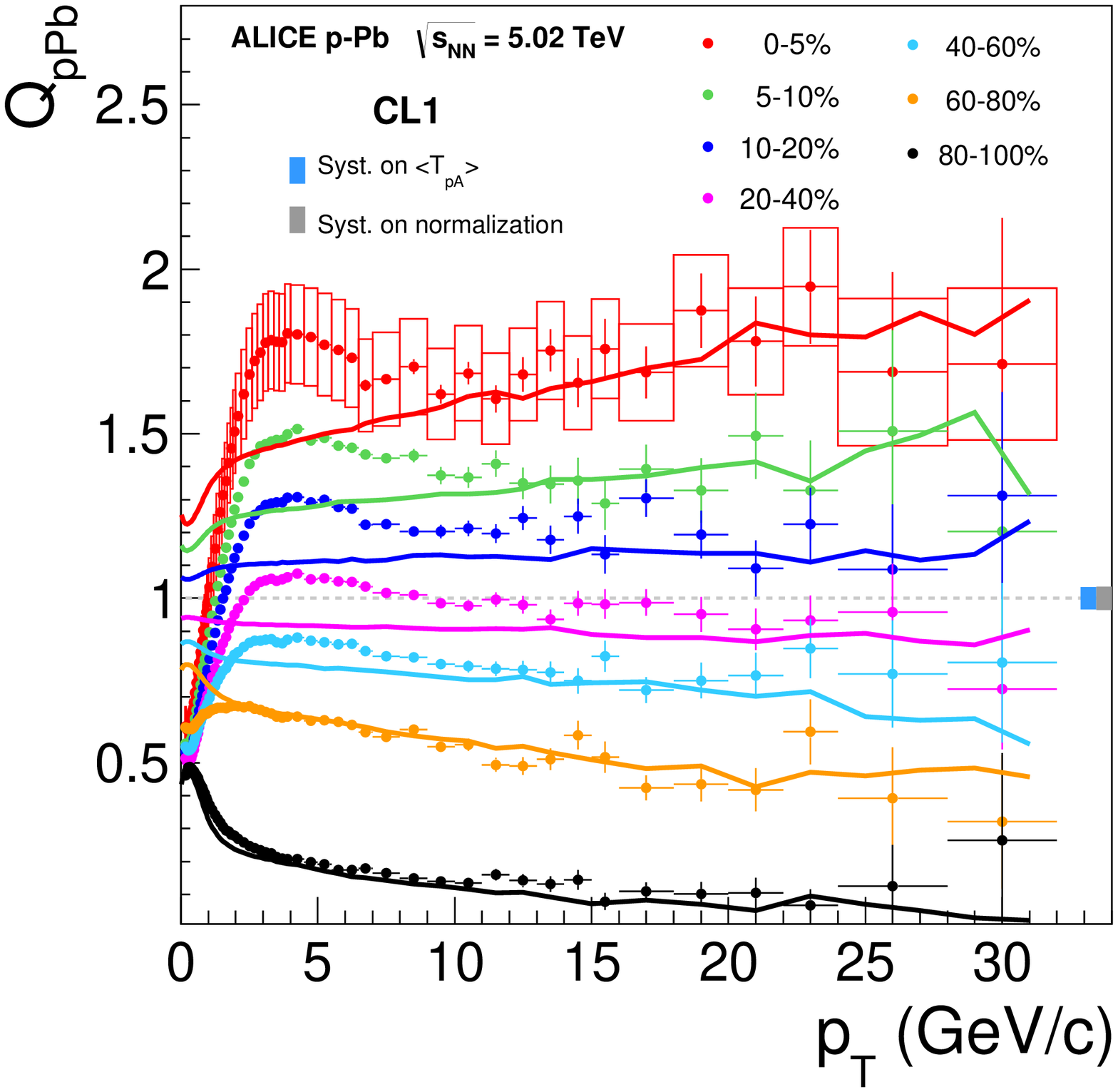}
\includegraphics[width=0.4\textwidth]{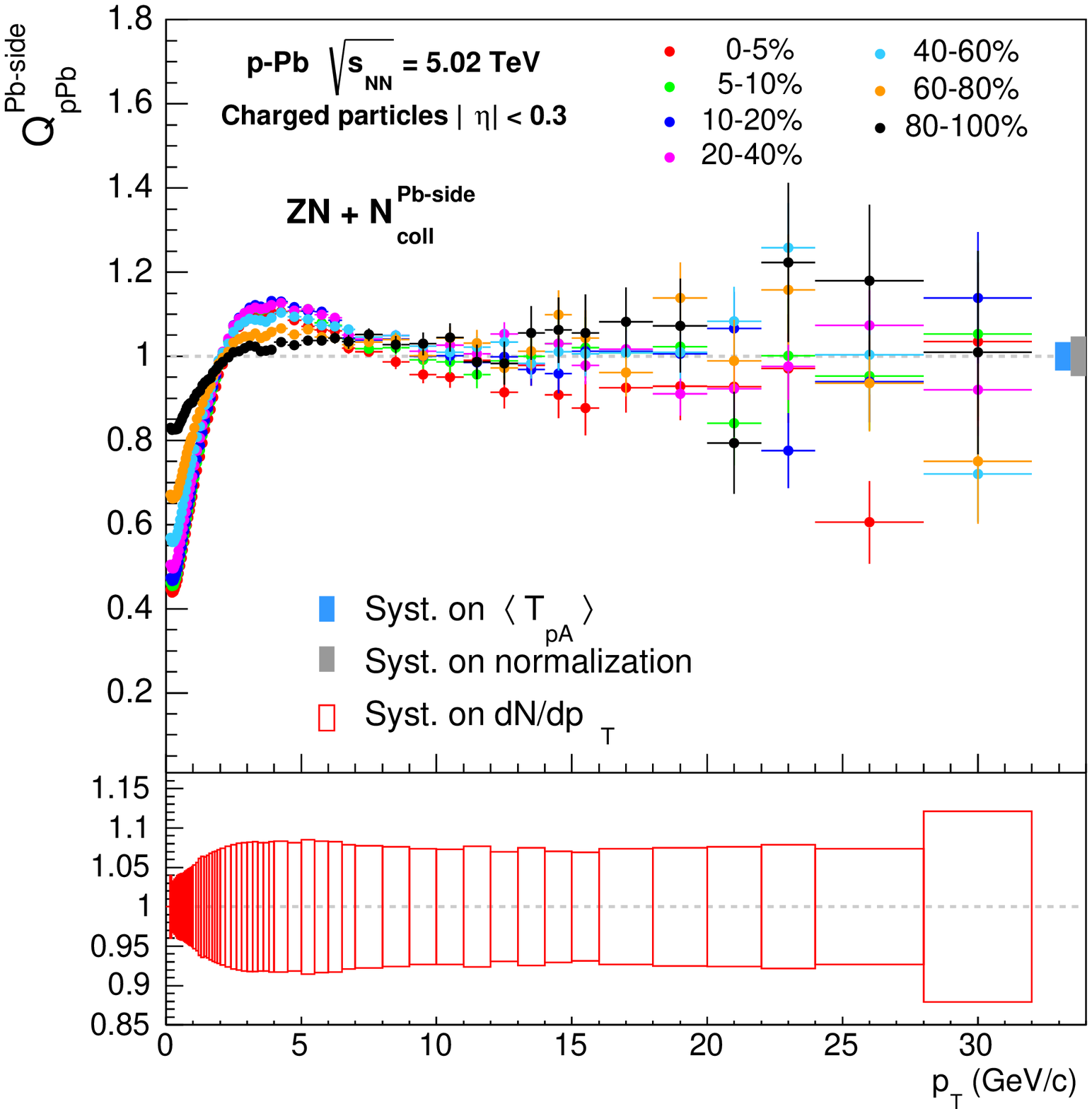}
 \caption{\qpa\ calculated with CL1 estimator (left), the lines are
   the G-PYTHIA calculations; with the hybrid method (right), spectra
   are measured in ZNA-classes and \Ncoll\ are obtained with the
   assumption that forward multiplicity is proportional to \Nparttar.
  \label{fig:QpAhybrid}}
\end{figure}

The nuclear modification factors \qpa\ calculated with
multiplicity based estimator (shown in Fig.\ref{fig:QpAhybrid} for
CL1, where centrality is based on the tracklets measured in
$|\eta|<1.4$) widely spreads between centrality classes. They also exhibit a
negative slope in \pT, mostly in peripheral events, due to the jet
veto bias, as jet contribution increases with \pT.  The \qpa\ compared
to G-PYTHIA, a toy MC which couples Pythia to a p-Pb Glauber MC, show
a good agreement, everywhere in 80-100\%, and in general at high-\pT,
demonstrating that the proper scaling for high-\pT\ particle
production is an incoherent superposition of \pp\ collisions, provided
that the bias introduced by the centrality selection is properly
taken into account, as eg in G-PYTHIA. 
With the hybrid method, using either the assumption on mid-rapidity
multiplicity proportional to \Npart, or forward multiplicity
proportional to \Nparttar, the \qpa\ shown in Fig.~\ref{fig:QpAhybrid},
are consistent with each other, and also consistent with unity for all
centrality classes, as observed for MB collisions, indicating the
absence of initial state effects. The observed Cronin enhancement is
stronger in central collisions and nearly absent in peripheral
collisions. The enhancement is also weaker at LHC energies compared to
RHIC energies.
\subsection{Charged particle density}
\begin{figure}[t!f]
 \centering 
 \includegraphics[width=0.4\textwidth]{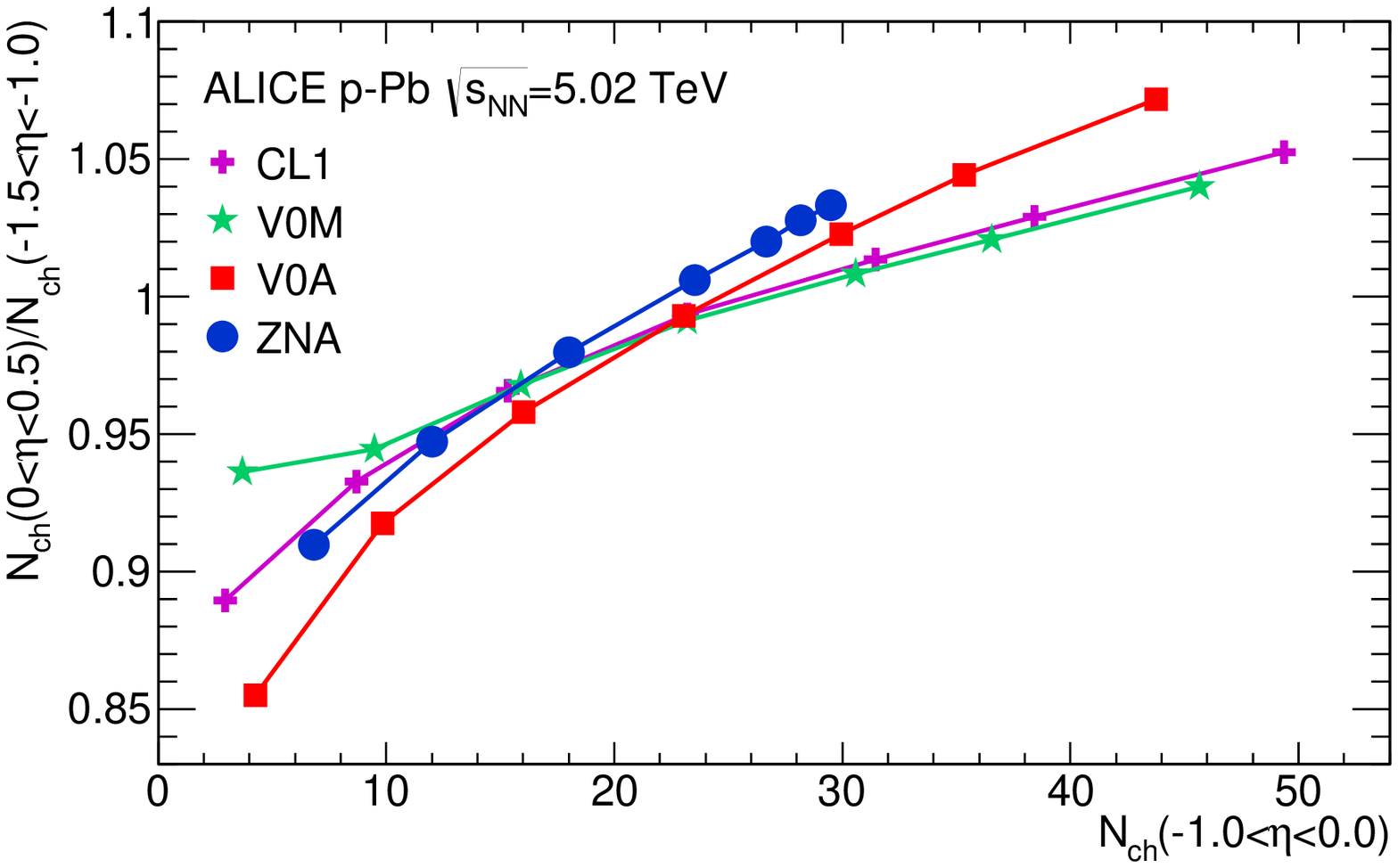}
 \includegraphics[width=0.3\textwidth]{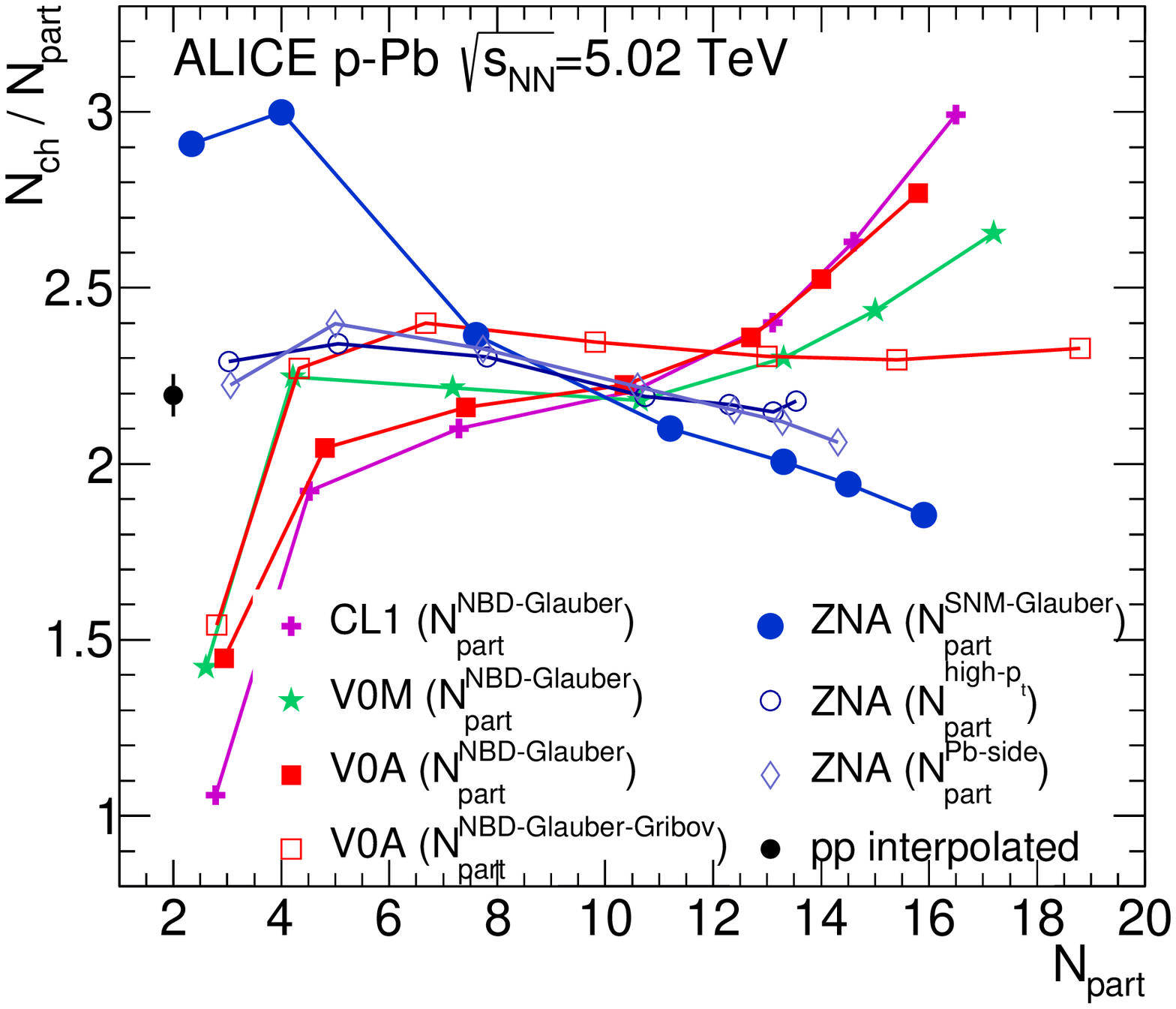}
 \caption{Left: Asymmetry of particle yield, as a function of the
   pseudorapidity density at mid-rapidity. Right: Pseudorapidity
   density of charged particles at mid-rapidity per participant as a
   function of \Npart\ for various centrality estimators.
\label{fig:dndetaNpart2}}
\end{figure}

Charged particle density is also studied as a function of $\eta$, for
different centrality classes, with different estimators. In peripheral
collisions the shape of the distribution is almost fully symmetric and
resembles what is seen in proton-proton collisions, while in central
collisions it becomes progressively more asymmetric, with an
increasing excess of particles produced in the direction of the Pb
beam. We have quantified the evolution plotting the asymmetry between
the proton and lead peak regions, as a function of the yield around
the midrapidity (see Fig.~\ref{fig:dndetaNpart2}, left): the
increase of the asymmetry is different for the different estimators.
Fig.~\ref{fig:dndetaNpart2} right shows \Nch\ at mid-rapidity divided
by \Npart\ as a function of \Npart\ for various centrality
estimators. For Multiplicity-based estimators (CL1, V0M, V0A) the
charged particle density at mid rapidity increases more than linearly,
as a consequence of the strong multiplicity bias. This trend is absent
when \Npart\ is calculated with the Glauber-Gribov model, which shows
a relatively constant behavior, with the exception of the most
peripheral point.  For ZNA, there is a clear sign of saturation above
\Npart\ = 10, due to the saturation of forward neutron emission. None
of these curves points towards the \pp\ data point. In contrast, the
results obtained with the hybrid method, using either
\Nparttar-scaling at forward rapidity or \Ncoll-scaling for
high-\pT\ particles give very similar trends, and show a nearly
perfect scaling with \Npart, which naturally reaches the
\pp\ point. This indicates the sensitivity of the \Npart-scaling
behavior to the Glauber modeling, and the importance of the
fluctuations of the nucleon-nucleon collisions themselves.

\section{Conclusions}
Multiplicity Estimators induce a bias on the hardness of the pN
collisions.  When using them to calculate centrality-dependent \qpa,
one must include the full dynamical bias.  However, using the
centrality from the ZNA estimator and our assumptions on particle
scaling, an approximate independence of the mid-rapitity multiplicity
on the number of participants is observed.  Furthermore, at
high-\pT\ the \pPb\ spectra are found to be consistent with the pp
spectra scaled by the number of binary collisions for all centrality
classes. Our findings put strong constraints on the description of
particle production in high-energy nuclear collisions.

\end{document}